\documentclass{aa}  

\usepackage{graphicx}
\usepackage{txfonts}
\usepackage[utf8]{inputenc}
\usepackage{amsmath}
\usepackage{amsfonts}
\usepackage{amssymb}
\usepackage{float}
\usepackage{graphicx}
\usepackage{subcaption}

\newcommand{\msunyr}{\ensuremath{\mathrm{M}_{\odot}{\rm yr}^{-1}}}   
\newcommand{\kms}{\ensuremath{{\rm km\,s^{-1}}}}                   
\newcommand{\msun}{\ensuremath{\mathrm{M}_{\odot}}}   
\newcommand{\msunano}{\ensuremath{\mathrm{M}_{\odot}{\rm yr}^{-1}}}  
\newcommand{\lsun}{\ensuremath{\mathit{L}_{\odot}}}                  
\newcommand{\rsun}{\ensuremath{\mathit{R}_{\odot}}}                  

\newcommand{\lstar}{\ensuremath{\mathit{L}_{\star}}}                 
\newcommand{\mdot}{\ensuremath{\dot{M}}}                             
\newcommand{\mstar}{\ensuremath{\mathit{M}_{\star}}}                 
\newcommand{\rstar}{\ensuremath{\mathit{R}_{\star}}}                 
\newcommand{\teff}{\ensuremath{\mathit{T}_{\rm eff}}}                
\newcommand{\vinf}{\ensuremath{v_{\infty}}}                          
\newcommand{\tstar}{\ensuremath{\mathit{T}_{\star}}}                 
\newcommand{\K}{\ensuremath{\mathrm{K}}}                 

\newcommand{\vcore}{\ensuremath{v_{\rm core}}}                         

\newcommand{\tauross}{\ensuremath{\tau_{\mathrm{Ross}}}}                 

\newcommand{\ang}{\ensuremath{\text{\AA}}}                  

\begin{document} 

\title{Catching a star before explosion: the luminous blue variable progenitor of SN~2015bh}
\author{I. Boian
          \inst{1}
          \and
          J. H. Groh\inst{1}
		}
   \institute{School of Physics, Trinity College Dublin, University of Dublin, Dublin, Ireland\\
              \email{boiani@tcd.ie , jose.groh@tcd.ie}
             }

     
  \abstract{
In this paper we analyse the pre-explosion spectrum of SN2015bh by performing radiative transfer simulations using the CMFGEN code. This object has attracted significant attention due to its remarkable similarity to SN2009ip in both its pre- and post-explosion behaviour. They seem to belong to a class of events for which the fate as a genuine core-collapse supernova or a non-terminal explosion is still under debate. Our CMFGEN models suggest that the progenitor of SN2015bh had an effective temperature between $8700$ and $10000$  K, luminosity in the range $ \simeq 1.8 ~-~4.74 \times 10^{6} ~ \lsun$, contained at least 25\% \ion{H}{} in mass at the surface, and half-solar \ion{Fe}{} abundances. The results also show that the progenitor of SN2015bh generated an extended wind with a mass-loss rate of $ \simeq 6 \times 10^{-4}$ to $ 1.5 \times 10^{-3} ~ \msunano$ and a velocity of $ 1000$ \kms. We determined that the wind extended to at least $ 2.57 \times 10^{14}$ cm and lasted for at least $30$ days prior to the observations, releasing $ 5 \times 10^{-5} ~ \msun$ into the circumstellar medium. In analogy to 2009ip, we propose that this is the material that the explosive ejecta could interact at late epochs, perhaps producing observable signatures that can be probed with future observations. We conclude that the progenitor of SN2015bh was most likely a warm luminous blue variable of at least $35 ~ \msun$ before the explosion. Considering the high wind velocity, we cannot exclude the possibility that the progenitor was a Wolf-Rayet star that inflated just  before the 2013 eruption, similar to HD5980 during its 1994 episode. If the star survived, late-time spectroscopy may reveal either a similar LBV or a Wolf-Rayet star, depending on the mass of the \ion{H}{} envelope before the explosion. If the star exploded as a genuine SN, 2015bh would be a remarkable case of a successful explosion after black-hole formation in a star with a possible minimum mass $35~\msun$ at the pre-SN stage.
    }
 
   \keywords{massive stars --
                supernovae --
                supernova impostors --
                radiative transfer --
                stellar evolution
               }
\maketitle

\section{Introduction}

Stars more massive than $8~\msun$ end their lives in violent events called core-collapse supernovae (SN). While SN explosions and their feedback are related to a multitude of topics in Astrophysics, the link between SNe and their progenitor stars is still not fully understood.

Massive stars eject large quantities of material into the interstellar medium throughout their evolution and before explosion \citep{dejager88,maeder_araa00,langer12,kiewe12,smith14araa,gme14}. Some stars undergo increased instabilities before death, leading to sudden and massive material ejections on short timescales of a few days to decades as shown, for example, by SN1988Z \citep{stathakis91,turatto93}, SN1994W \citep{sollerman98,chugai04}, SN1998S \citep{leonard00,fassia01}, SN2005gj \citep{trundle08}, SN 2005gl \citep{galyam07,galyam09}, and SN2006gy \citep{smith07}. This material builds up around the star forming a circumstellar medium (CSM). The SN ejecta crashes into this medium and the kinetic energy of this interaction is partially converted into radiation. Depending on the CSM density, the interaction can give rise to luminous transients comparable to the luminosities of the actual SN \citep[e.g.,][]{chevalier94}. Due to the radiation originating in the slow-moving CSM (typically $100-1000$ \kms), the spectrum of these events is characterised by relatively narrow emission lines \citep{chugai01,dessart09,dessart15}. Therefore these explosions, known as interacting SNe or SN IIn (\citealt{schlegel90}), can be used as a tool to probe the circumstellar material and constrain properties of the progenitor such as abundances, mass-loss rates, wind velocities and radius \citep{groh14,grafener16}.

Depending on the density and extension of the CSM, progenitor properties can be directly retrieved from follow-up spectroscopic observations obtained from a few hours up to decades after the explosion. For example, SN2013fs shows signatures of CSM interaction for the first 2 days and afterwards it behaves as a regular SN IIP (\citealt{yaron17}). In contrast, the interaction can last for many years when the material around the star is dense and extended, as in the case of SN 1998S \citep{mauerhan12b,shivvers15}. There is mounting evidence of SNe that are intermediate to compact, relatively low-dense CSMs of SN IIP and extended, dense CSM of SN IIn. Examples of intermediate SNe are SN 2013cu (IIb; \citealt{galyam14,groh14}) and iPTF13iqb \citep{smith15}.

For some events, the continuing interaction makes it difficult to determine whether it is a genuine SN that interacts with a pre-existing CSM or if the star survived after a series of massive eruptions, in which case it is deemed a SN impostor \citep[e.g.,][]{vandyk00}. One of the best observed transients for which the fate is under discussion is SN2009ip, which has been monitored photometrically and spectroscopically with exquisite time coverage \citep{fraser13,levesque14,margutti14,mauerhan13,ofek13_2009ip,pastorello13,prieto13,smith14}.The spectrum, energetics, low \ion{Ni}{} mass and late time evolution resemble both SN and the SN impostors scenarios, with \cite{graham17}, \cite{smith14}, \cite{ofek13_2009ip}, \cite{mauerhan13} favouring the core-collapse case, while \cite{fraser15} states that there is no conclusive evidence for it. \cite{moriya15} proposes a two-component medium, an inner shell and an outer wind. Inside this medium, an explosion took place and the material crashed into the inner shell creating the apparent SN. The slow decay is maintained by the continuous interaction with the outer wind. The explosion properties estimated by \cite{moriya15} point to either a SN impostor or a peculiar SN. SN2009ip is still showing signs of interaction, masking the remnant of the event, if any.

Recent studies suggest that several other events are similar to 2009ip, forming a distinct class of transients \citep{thone16,eliasrosa16,pastorello18}. Among these, SN2015bh bares remarkable resemblance to SN2009ip, as discussed extensively in \cite{thone16}, \cite{eliasrosa16}, \cite{pastorello18}, and has a unique pre-explosion spectrum observed at the lowest ever flux level for this class of transients. SN2015bh, also known as SNhunt275 or PTF13efv, was first officially reported as a SN candidate in February 2015 (\citealt{eliasrosa15}), due to a brightening event dubbed 2015A that peaked at an absolute magnitude of $M_\mathrm{R} = -15$ mag. This was the first of two lightcurve peaks, with the second peak, now labelled 2015B, reaching $M_\mathrm{R} = -17.5$ mag on May 2015.

There is a significant amount of information on the precursor of SN2015bh. Its host galaxy was monitored for 21 years before the discovery of SN2015bh, which provided extended photometric data of the progenitor. This revealed a long-term photometric variation of $ \pm 2$ mag sometimes showing sudden changes in brightness in 2008, 2009 and 2013 (\citealt{thone16}). Serendipitous spectroscopy of the progenitor of SN2015bh was obtained on 12 November 2013, when its  absolute magnitude was $ M_\mathrm{R} \sim -10.5$ mag. Although the star has documented variability, at this point the spectrum does not show any signs of interaction. This spectrum is characterized by \ion{H}{} and \ion{Fe}{ii} emission \citep{thone16}, and is one of the first times that a spectrum of a potential SN progenitor has been observed. Therefore, these observations reveal invaluable information on massive stars and their pre-SN behaviour. The 12 November 2013 observations were taken right at the onset of another eruption, detected in December 2013 by iPTF (\citealt{ofek16}). The first iPTF photometric detection of this outburst was on 26 November 2013, but \cite{ofek16} mention the possibility that it had started at an earlier date. After the December 2013 event, SN2015bh shows another sudden increase in brightness of $2$ mag. The light curve of SN2015bh is discussed in detail in both \cite{thone16} and \cite{eliasrosa16}.

\cite{ofek16} analyse the pre-explosion spectrum of SN2015bh obtained during the December 2013 outburst. The spectrum exhibits a strong, narrow $H\alpha$ line with a P-Cygni absorption component extending up to $-1300$ \kms. By fitting a blackbody to the continuum, \cite{ofek16} obtained an effective temperature of $5750$ K and a radius of $4 \times 10^{14}$ cm. Assuming a super-Eddington continuum-driven wind, they estimated the total mass lost during the December 2013 event to be $ M \simeq 4 \times 10^{-5} \msun $.
\cite{thone16} analyse both the pre and post-explosion observations, proposing that SN2015bh was a luminous blue variable (LBV) star in outburst for over 20 years that experienced a possible SN explosion (2015B) after several precursor events. They also entertain the possibility of the star surviving and becoming a WR star and, given the similarities to a number of other events (SN2009ip, SNhunt248, and SN1961V) they propose the existence of a new category of transients. The members of this category show variations of $\simeq 2$ mag for at least a few decades, a bright precursor rapidly followed by a main event resembling a SN, and LBV-type spectra during the outbursts and until after the maximum of the main event. \cite{eliasrosa16} also propose that SN2015bh was a massive blue star. However, they argue that the SN explosion was actually the 2015 A event, explaining the low luminosity as due to massive fallback of material onto a collapsed core. They further suggest that the 2015B event was the result of the SN ejecta interacting with a dense CSM. \cite{eliasrosa16} do not completely rule out a non-terminal explosion either.

SN2015bh offers the unique opportunity of investigating the pre-SN spectrum of a massive star before explosion. This paper is organised as follows. In Section 2 we present the characteristics of CMFGEN, the code that was used to model the spectrum of the SN2015bh progenitor. Section 3 provides a detailed discussion on the results obtained from our models, while Section 4 puts the results in the context of other SNe, SN impostors and other similar events. The main points of the paper are summarised in Section 5.

\section{Radiative Transfer Modelling}
We investigate the properties of the SN2015bh progenitor by computing radiative transfer models of the outflow produced by the star. We fit the spectrum obtained on 12 November 2013 by \cite{thone16} with the Gran Telescopio Canarias (GTC) and downloaded via the WISeREP\footnotemark{} repository \citep{yaron12}. We employ the line-blanketed atmospheric/wind radiative transfer code CMFGEN (\citealt{hm98}). The code computes continuum and line formation in non-local thermodynamic equilibrium and spherical symmetry, and it does not account for time-dependent effects. Including clumping effects in the winds of massive stars has improved the modelling of spectral observations in many cases, such as Eta Car \citep{hillier01,ghm12}, AG Car \citep{ghd09}, the Pistol Star and FMM362 \citep{najarro09}. Our models include clumping with volume filling factor of $f=0.1$,  and we explore the effects of clumping in Sect. 3.3.

\footnotetext{https://wiserep.weizmann.ac.il}

\begin{figure}
\includegraphics[width=1.02\columnwidth]{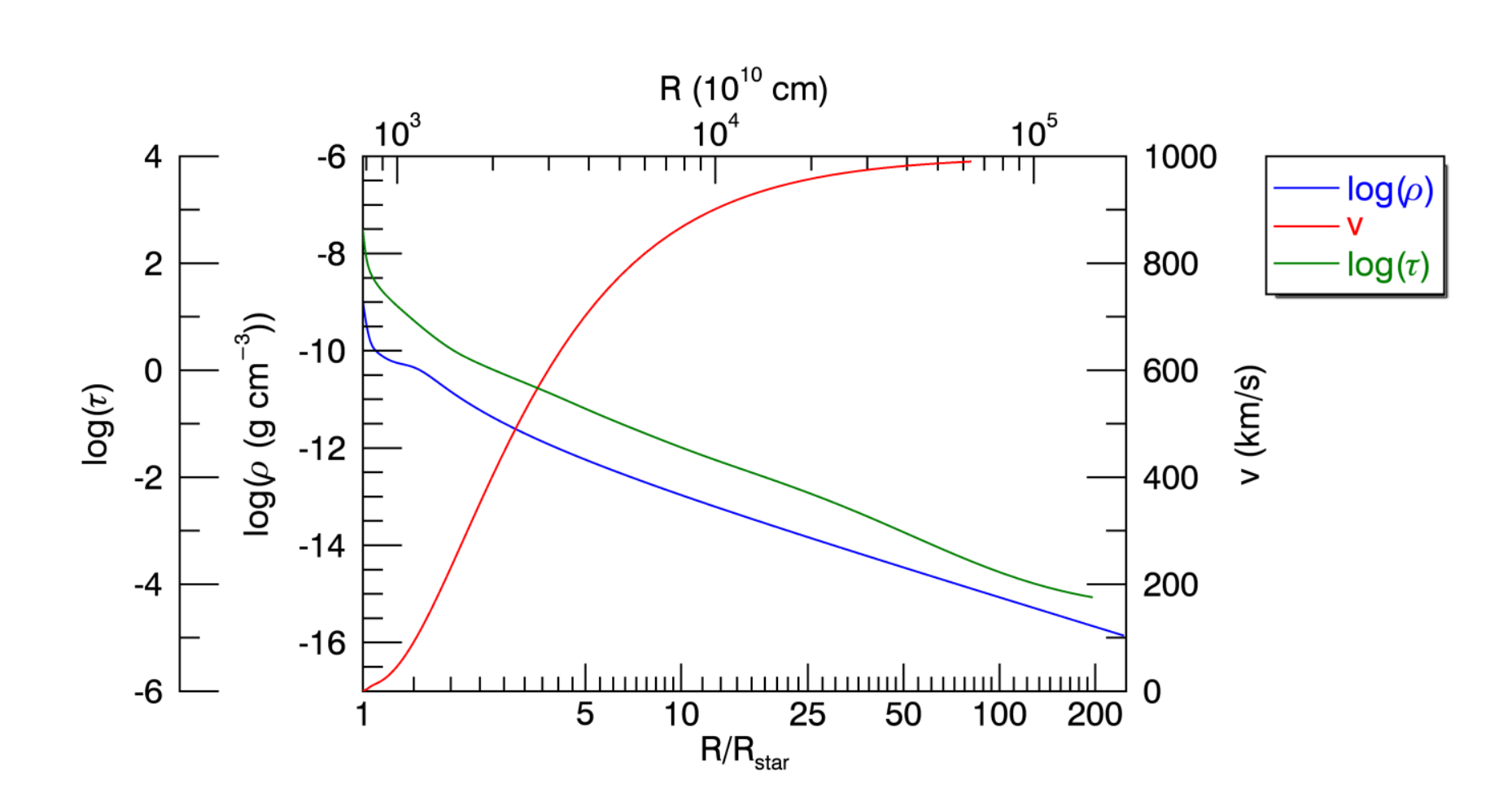}
\caption{ The velocity (red), density (blue), and Rosseland optical depth structures (green) for the 2015bh progenitor based on the CMFGEN modelling of the 12 November 2013 spectrum.}
\label{velden}
\end{figure}

The input physical quantities that the code requires to compute the spectrum are the inner boundary radius ($R_{\star}$), a constant mass loss rate ($\dot{M}$), the wind terminal velocity (\vinf),  abundances, and the bolometric luminosity (\lstar). CMFGEN does not solve for a self-consistent hydrodynamical solution for the outflow. Instead we assume a density scale height for the atmosphere of the star of $0.02 ~ \rstar$ joined smoothly to the wind just below the sonic point (at $v = 8 $ \kms). The stationary wind has a density profile derived from the mass continuity equation:
\begin{equation}
\rho (r) = \frac{\dot{M}}{4 \pi r^2 v(r)}
\end{equation}
and accelerates following a beta-type law: 
\begin{equation}
v(r)=\frac{v_0+(v_{\infty}-v_{0})(1-R_{\star}/r)^{\beta}}{(1+v_0/\vcore \exp{([R_{\star}-r]/h_\mathrm{eff})}}.
\end{equation}
where $v_o$ is the velocity at a reference radius close to $\rstar$, \vcore\ is the velocity at the inner radius $\rstar$, $\vinf$ is the terminal velocity of the wind and $h_\mathrm{eff}$ is the scale height in units of $R_{\star}$. Our models assume $v_0 = 10$ \kms, $\vcore= 0.85$ \kms, $h_\mathrm{eff} = 0.02 \rstar$ and $\beta = 2.5$.  The $\beta$ parameter describes the steepness of the velocity law and we have chosen a value typical of LBVs. For example, other similar works that have led to this $\beta$ value for LBVs include \cite{najarro97}, \cite{najarro01} for P Cygni, \cite{najarro09} for the Pistol Star, and \cite{hillier98} for HDE 316285. We discuss the effects of $\beta$ in Sect. 3.3. Figure \ref{velden} shows the velocity, density, and optical depth structures of SN 2015bh's progenitor.

We assume diffusion approximation at the inner boundary and adjust \vcore\ to obtain a Rosseland optical depth of \tauross=150 at \rstar. Because the 2015bh progenitor has a dense wind, the photosphere is located in the wind. Hereafter we define \teff\ as the effective temperature computed at the radius where \tauross=2/3, and likewise \tstar\ at  \tauross=20. The outer boundary is defined at $R_\mathrm{out}= 250 ~ \rstar$ and is chosen to be large enough to account for all the emission line region. The region between $\rstar$ and $R_\mathrm{out}$ is split into $77$ depth points where we calculate the values of each physical quantity necessary to simultaneously solve for the energy level populations and the properties of the radiation field. The atoms included in the model are summarised in Table 1, together with the number of 'super' levels\footnotemark{}  used and the number of atomic levels.  Our models account for the effects of charge-exchange reactions. In the context of this paper, one reaction of major importance for the \ion{Fe}{ii} spectrum is $\mathrm{Fe^{2+}+H \leftrightarrow Fe^{+}+H^{+}}$ \citep{hillier01}, since it affects the \ion{Fe}{} ionization structure. The atomic data is provided by a number of sources, as described in detail in \cite{hm98}.

\footnotetext{The inclusion of 'super' levels is a technique used to decrease the number of levels whose atomic populations must be explicitly solved for (\citealt{anderson89,hm98}), thus simplifying the calculations and improving computational time and memory requirements.} 

After computing the atmospheric and wind structure, we use CMF\_FLUX (\citealt{bh05}) to calculate the synthetic spectrum in the observer's frame. We determine the properties of the SN2015bh progenitor mainly by comparing the observed optical spectrum ($4000 - 6700$ \ang) to the synthetic spectrum generated from our models. While the original model spectra has an arbitrarily high spectral resolution, we use a degraded version by convolving the original model spectrum with a Gaussian function with full-width at half maximum FWHM = $300$ \kms to better match the spectral resolution of the observations. Fitting the strength of the emission lines to the ones in the observed spectrum, we were able to constrain $\mdot$, $\teff$ indirectly by modifying $\rstar$ and $\lstar$, and the metallicity $Z$. Then, we matched the width of the emission lines to obtain a good approximation for $\vinf$. The value of $\lstar$ is constrained by comparing the observed flux to our model and accounting for interstellar extinction using the reddening law from \citet{fitzpatrick99}. \cite{thone16} quote a distance of $d=27~\mathrm{Mpc}$ to the host galaxy, while \cite{eliasrosa16} derived $d=29.3 \pm 2.1 ~\mathrm{Mpc}$ from the recessional velocity of the galaxy. We have added the distance uncertainty to our luminosity calculations. Following this procedure, we found a best-fit model that reproduces the observed properties reasonably well and is discussed in detail in the next section. 

\begin{figure*}
\includegraphics[scale=0.6]{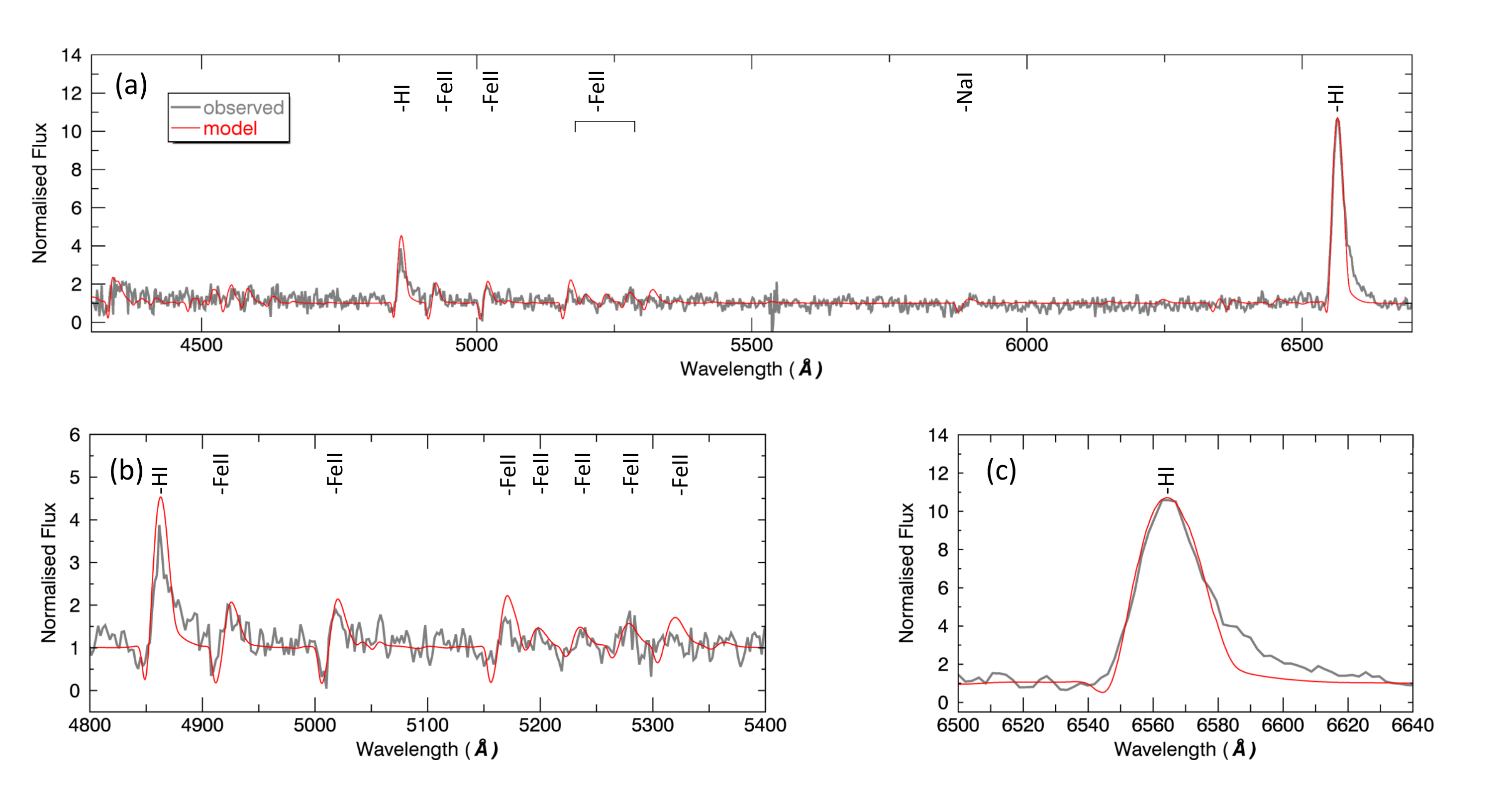}
\caption{Panel (a): Optical spectrum of one of our best-fitting models (red) compared to the observed spectrum taken on 12 November 2013 (gray) in the 4300--6700~\ang\ region. See text for model parameters. Panel (b): Zoom-in from 4800 \ang to 5400 \ang, containing the H$\beta$ line and a forest of \ion{Fe}{ii} lines. Panel (c): Zoom-in around the H$\alpha$ line. }
\label{modelfit}
\end{figure*}

\begin{figure*}
\includegraphics[scale=0.55]{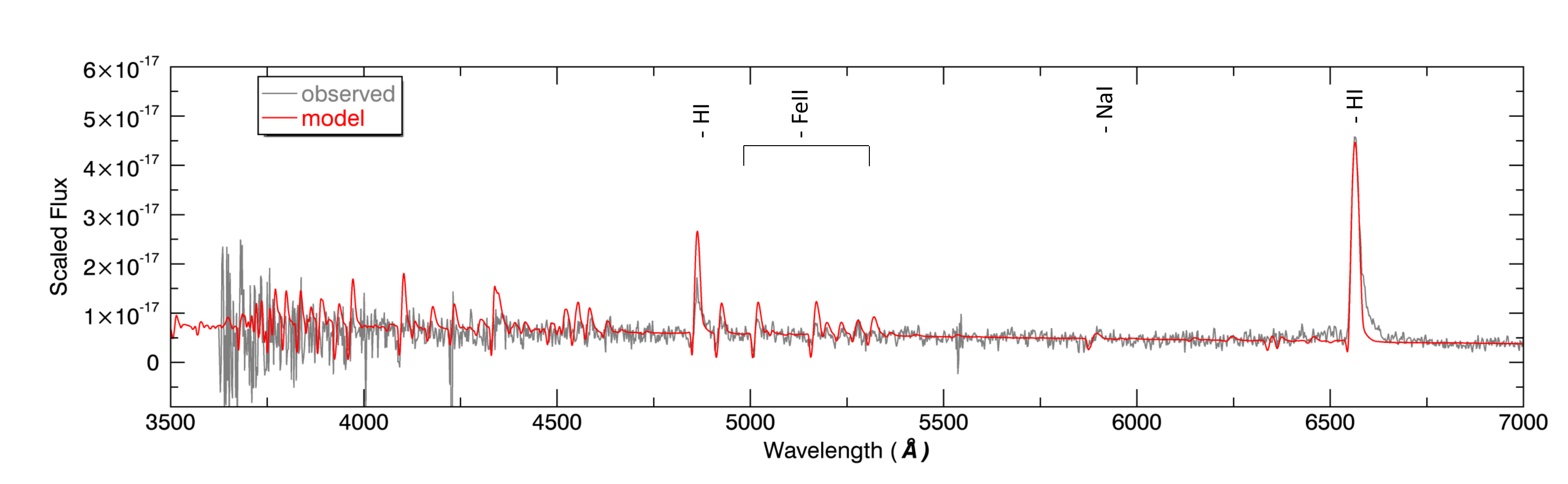}
\caption{Flux-calibrated spectrum of SN2015bh observed on 12 November 2013 (gray) compared to one of our best-fit CMFGEN models (red). This model has been scaled to a distance of $d=27~\mathrm{Mpc} $ and reddened using E(B-V)$=0.25$ and R$_\mathrm{V}=3.1$.}
\label{modelfitflux}
\end{figure*}

\begin{table}
\centering
\begin{tabular}{c c c}
\hline
\hline
\multicolumn{3}{c}{Summary of Model Atoms} \\
\hline 
Species & No. of super-levels & No. of atomic levels \\
\hline
\ion{H}{i} & 20 & 30 \\
\ion{He}{i} & 40 & 45 \\
\ion{He}{ii} & 22 & 30 \\
\ion{C}{i} & 38 & 80 \\
\ion{C}{ii} & 39 & 88 \\
\ion{C}{iii} & 32 & 59 \\
\ion{N}{i} & 44 & 104 \\
\ion{N}{ii} & 157 & 442 \\
\ion{N}{iii} & 42 & 158 \\
\ion{O}{i} & 69 & 161 \\
\ion{O}{ii} & 26 & 80 \\
\ion{O}{iii} & 33 & 92 \\
\ion{Na}{i} & 18 & 44 \\
\ion{Mg}{i} & 37 & 57 \\
\ion{Mg}{ii} &18 & 45 \\
\ion{Al}{ii} & 38 & 58 \\
\ion{Al}{iii} & 17 & 45 \\
\ion{Si}{ii} & 22 & 43 \\
\ion{Si}{iii} & 20 & 34 \\
\ion{Si}{iv} & 22 & 33 \\
\ion{Ca}{i} & 23 & 39 \\
\ion{Ca}{ii} & 17 & 46 \\
\ion{Ti}{ii} & 33 & 314 \\
\ion{Ti}{iii} & 33 & 380 \\
\ion{Fe}{i} & 69 & 214 \\
\ion{Fe}{ii} & 67 & 403 \\
\ion{Fe}{iii} & 48 & 346 \\
\ion{Ni}{ii} & 29 & 204 \\
\ion{Ni}{iii} & 28 & 220 \\
\hline                                                                                                                                                                                                                                                                                                                                                                                                                                                                                                                                                                                                                                                                                                                                                                                                                                                                                                                                                                                                                                                                                                                                                                                                                                                                                                                                                                                                                                                                                                                                                                                                                                                                                                                                                                                                                                                                                                                                                                                                                                                                                                                                                                                                                                                                                                                                                                                                                                                                                                                                                                                                                                                                          
\end{tabular}
\label{atomic}
\caption{The atomic model used in the CMFGEN analysis of 2015bh.}
\end{table}

\section{Results}

\subsection{Physical Properties of SN2015bh's Progenitor \label{res.properties}}

The SN2015bh spectrum taken on 12 November 2013 shows strong \ion{H}{i} $ \lambda6363 ~ \ang$ emission, a multitude of \ion{Fe}{ii} $ \lambda \lambda 4924, 5018, 5196, 5198, 5235, 5276, 5317 ~ \ang $ lines and a \ion{Na}{i} $\lambda 5889 ~ \ang$ line. All the lines have P-Cygni profiles. Figure \ref{modelfit} shows the observed normalised 2013 spectrum of SN2015bh (gray) and the normalised synthetic spectrum from one of our best-fitting models (red) at optical wavelengths. Our models  reproduce the strength of the H$\alpha$ line  reasonably well (Fig.~\ref{modelfit}c), but it slightly overestimates the \ion{H}{$\beta $} emission (Fig.~\ref{modelfit}b). This could be due to the ionization and/or density structures not being fully reproduced by our models. In a similar manner, most \ion{Fe}{ii} lines are well fitted, but two of them are overestimated (\ion{Fe}{ii} $\lambda 5169 ~ \ang$ and \ion{Fe}{ii} $\lambda 5317 ~ \ang$). A summary of the physical properties of the progenitor of SN2015bh is presented in Table~\ref{modelparam}.

We can constrain $\mdot$ and $\tstar$ by fitting the strength of the \ion{H}{i} and \ion{Fe}{ii} lines. For the model in Fig. \ref{modelfit} we have $\mdot = 10^{-3} ~\msunano$ and $\tstar = 15000$ K. Because of the high wind density, the optical depth towards the hydrostatic layers of the star is  $>> 1$. The photosphere is extended and formed in moving layers, in a similar way as other LBVs such as AG Car \citep{ghd09,ghd11}, HR Car \citep{gdh09}, P Cygni \citep{najarro97,najarro01} and Eta Car \citep{hillier01,ghm12}. This causes \teff\ to be lower than \tstar\ (see discussion in \citealt{ghd09}), and our CMFGEN model shown in Fig. \ref{modelfit} has $\teff=8700~\K$.  Other combinations of $\mdot$ and $\tstar$ would also fit due to degeneracy, which is further analyzed in Section 3.3. Table 2 lists the ranges of possible values for the parameters of SN2015bh's progenitor. 

Figure \ref{modelfitflux} displays the flux-calibrated observed spectrum of SN2015bh in 12 November 2013 and one of our best fit models, assuming $R_\mathrm{V}=3.1$ and $d=27 Mpc$. By comparing the absolute flux levels, we are able the constrain \lstar\ and the color excess E(B-V). Under these assumptions, our CMFGEN models indicate that the bolometric luminosity of SN2015bh on 12 November 2013 is $\lstar=2.7\times10^6~\lsun$ and E(B-V)=0.25. The color excess is in remarkable agreement with the value estimated by \citet{thone16} based on the strength of interstellar Na lines (E(B-V)=0.21). However, we were also able to fit the optical spectrum using other values of $R_\mathrm{V}$ between $2$ and $5$. The differences in $R_\mathrm{V}$ have more influence at smaller wavelengths, therefore we would like to stress the importance of multi-wavelength observations for future events. Taking under consideration the uncertainties in $R_\mathrm{V}$ and $d$, the $\lstar$ ranges from $1.8 \times 10^{6} \lsun$ ($\log(\frac{\lstar}{\lsun})=6.25$) for the lowest values of $R_\mathrm{V}$ and $d$, to $4.74 \times 10^{6} \lsun$ ($\log(\frac{\lstar}{\lsun})=6.67$) for the highest values of $R_\mathrm{V}$ and $d$. 

The morphology and width of the lines allowed us to determine $\vinf$, which we have estimated to be $ \simeq 1000 $ \kms. The line profiles also show an asymmetric bump on the left side of the \ion{H}{$\alpha$} and \ion{H}{$\beta$} lines. This could indicate a larger wind velocity, but we discuss various possibilities in Sect. 3.3.

We have used the strength of the \ion{Fe}{ii} lines to determine the \ion{Fe}{} abundance of the SN2015bh progenitor. Our models indicate the progenitor star had about half-solar \ion{Fe}{} abundance. It is interesting to contrast the \ion{Fe}{} abundance of SN2015bh to the \ion{O}{} abundance of the surrounding regions of SN2015bh. \cite{thone16} suggest that the \ion{O}{} abundance of the surrounding regions is half-solar. In principle, there is no reason to expect that the progenitor of SN2015bh will have the same \ion{O}{} abundance as that of its environment. In particular for LBVs, the \ion{O}{} abundance is expected to be severely affected by the presence of CNO-burning products at the surface \citep{ghd09,gme14}.  Unfortunately, the progenitor spectrum analysed here does not show any strong emission of CNO lines, and therefore the abundance of these elements cannot be constrained using solely the 12 November 2013 spectrum.

The model discussed in this section has $X=0.49$ for \ion{H}{} and $Y=0.50$ for \ion{He}{}. However, we have found that, for different values of $\mdot$ and $\tstar$, we can still reasonably fit the observed spectrum for \ion{H}{} abundances in the range $X=0.25$ to $X=0.75$. A similar degeneracy has been found for other LBVs, such as HDE 316285 \citep{hillier98}. Throughout the range of parameters of our models, the \ion{Na}{i} $ \lambda 5889 ~ \ang$ line showed the weakest dependence on $\mdot$ and $\tstar$. However, the morphology of the \ion{Na}{i} $ \lambda 5889 ~ \ang$ line is affected by  \ion{He}{i} $ \lambda 5875 ~ \ang$ emission, which appears when the model has low $\mdot$ or high $\tstar$.

We can estimate a lower limit for the mass of the progenitor given the luminosity obtained from the modelling. By assuming that the star is at the Eddington limit, we have $\lstar=L_\mathrm{EDD}$. The relation between $L_\mathrm{EDD}$ and the mass required for stability is:

\begin{equation}
L_\mathrm{EDD} = 3.22 \times 10^4 \Big[ \frac{N(H)/N(He)+4}{N(H)/N(He)+2} \Big] \Big( \frac{M}{\msun} \Big) \lsun\,.
\end{equation}

Considering the assumptions discussed previously, our best-fit model indicates that $ \lstar = 2.7 \times 10^6 \lsun$ and that $N(H)/N(He) = 3.846$, therefore the minimum mass of the progenitor of SN2015bh at the pre-explosion stage is $M_\mathrm{min} \simeq 62~\msun$. However, taking into account the \ion{H}{} abundance and the $\lstar$ uncertainties, the minimum mass range changes significantly, giving $M_\mathrm{min}=35-120~\msun$.

\begin{table}[h!]
\centering
\begin{tabular}{c c c}
\hline
\hline
\multicolumn{3}{c}{SN2015bh Progenitor Model Parameters} \\
\hline\multicolumn{1}{c}{\lstar } & \multicolumn{2}{l}{$ 1.8 ~-~4.7 \times 10^{6} ~ \lsun$} \\
\multicolumn{1}{c}{\rstar } & \multicolumn{2}{l}{$ 5.1 ~ - ~ 25.9 \times 10^{12}$\,cm (74--374~\rsun)} \\
\multicolumn{1}{c}{\vinf } & \multicolumn{2}{l}{$1000 $ \kms } \\
\multicolumn{1}{c}{\mdot } & \multicolumn{2}{l}{$0.6~ -  ~1.5 \times 10^{-3}~\msunano$} \\
\multicolumn{1}{c}{\tstar} & \multicolumn{2}{l}{$13000 - ~19500 $ K } \\
\multicolumn{1}{c}{\teff} & \multicolumn{2}{l}{$8700 - ~10000 $ K } \\
\multicolumn{1}{c}{$E(B-V)$ } & \multicolumn{2}{l}{$0.25$ } \\
\hline
\multicolumn{3}{l}{Abundances} \\
\hline
Element & Mass Fraction ($\chi$) & $c/c_{Sun}$ \\
\hline
H & $0.49$ & $0.7$ \\
He  & $0.5$ & $1.8$ \\
C  (assumed) & $5.6 \times 10^{-5}$ & $0.02$  \\
N  (assumed) & $8.2 \times 10^{-3}$ & $7.5$  \\
O  (assumed) & $1.3 \times 10^{-4}$ & $0.02$  \\
Na  & $3.5 \times 10^{-5}$ & $1$  \\
Si   (assumed) & $6.6 \times 10^{-4}$ & $0.9$ \\
Fe  & $6.8 \times 10^{-4}$ & $0.5$ \\
\hline
\end{tabular}
\caption{The CMFGEN model parameter ranges for the best-fitting models of SN2015bh's progenitor. The abundances correspond to those from the model shown in Figure 2.  The CNO and Si abundances have been assumed, while the H/He abundance is degenerate with \mdot. See text for details.}
\label{modelparam}
\end{table}

\subsection{CSM Properties}

\begin{figure}
\centering
\includegraphics[scale=0.5]{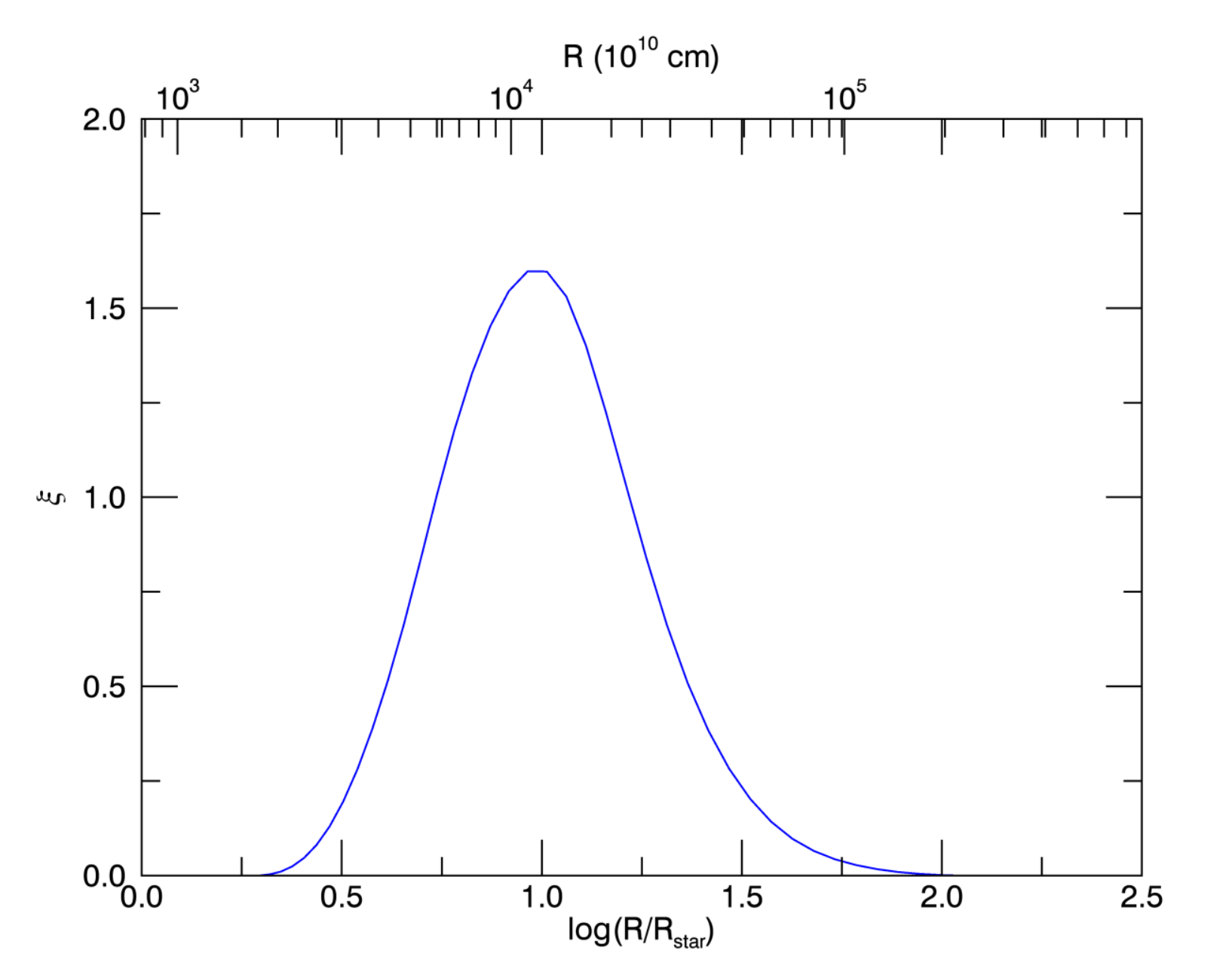}
\caption{Formation region of the H$\alpha$ line as a function of distance to the star. The quantity $\xi$ is related to the equivalent width (EW) of the line as $EW=\int_{\rstar}^{\infty} \xi (R) d(\log(R))$.}
\label{halphalineform}
\end{figure}

Based on our CMFGEN models, we can infer lower limits for the properties of the CSM surrounding the progenitor of SN2015bh, such as its extent and mass, and the duration of the stellar wind ejection required to give the November 2013 spectrum.

-- {\it CSM extension:} CMFGEN is able to compute the regions where the spectral lines are formed, which allows us to estimate how extended the CSM is. We use for this purpose the H$\alpha$ line since this is the line that we predict to form over the largest distance from the star (Fig. \ref{halphalineform}). The extension of the \ion{H}{$\alpha$} emitting region is estimated as $r_\mathrm{CSM}=r_{H\alpha}=10^{1.7} ~ \rstar = 50 ~ \rstar \simeq 2.57 \times 10^{14}$ cm. This is a lower limit for the extension of the CSM, since we do not have other optical diagnostics formed over larger distances. 

-- {\it Duration of stellar wind ejection:} We can determine the minimum time, $\Delta t$, required for an outflow of $v=1000 $ \kms to extended to $r_\mathrm{CSM} = 2.57 \times 10^{14}$ cm as\begin{equation}
\Delta t \simeq  r_\mathrm{CSM} \times v^{-1} \simeq  30 ~ \mathrm{days} \Big( \frac{r}{2.57 \times 10^{14} \mathrm{cm}} \Big) \Big( \frac{v}{1000 \kms} \Big)^{-1}\, . 
\end{equation}

-- {\it CSM mass: } Another quantity we determined using the CMFGEN models is the mass loss rate. {The lower limit for the mass loss rate from our models is $\mdot = 0.6 \times 10^{-3} ~\msunano$. Given that the stellar wind ejection lasted at least $ \Delta t = 30$ days, the amount of material ejected by the star in the surrounding medium in $\Delta t$ is
\begin{equation}
M_\mathrm{CSM} \simeq 5 \times 10^{-5} \msun \Big( \frac{\Delta t}{30 \mathrm{days}} \Big)\,.
\end{equation}

-- {\it CSM energetics:} Our CMFGEN models indicate a minimum radiative luminosity of $L_\mathrm{RAD} = 1.8 \times 10^6 ~ \lsun$ and kinetic luminosity of $L_\mathrm{KIN} = 1.2186 \times 10^4 ~ \lsun$. Assuming a duration of the stellar wind ejection of 30 days, this corresponds to a radiated energy of $E_\mathrm{RAD} \simeq 1.8 \times 10^{46}~$erg and kinetic energy of $E_\mathrm{KIN} = 1.26 \times 10^{44}$~ erg. This leads to a radiative efficiency of $ \epsilon = E_\mathrm{RAD}/ E_\mathrm{KIN} \simeq \mathrm{140}$, which is within the range expected for stellar winds of hot massive stars.

\subsection{Sensitivity of the Derived Parameters and Model Degeneracies}
In this section we explore possible degeneracies of our best-fit model by analysing a grid of synthetic CMFGEN spectra covering a large parameter space.

\begin{figure*}
\centering
\includegraphics[width=0.89\textwidth]{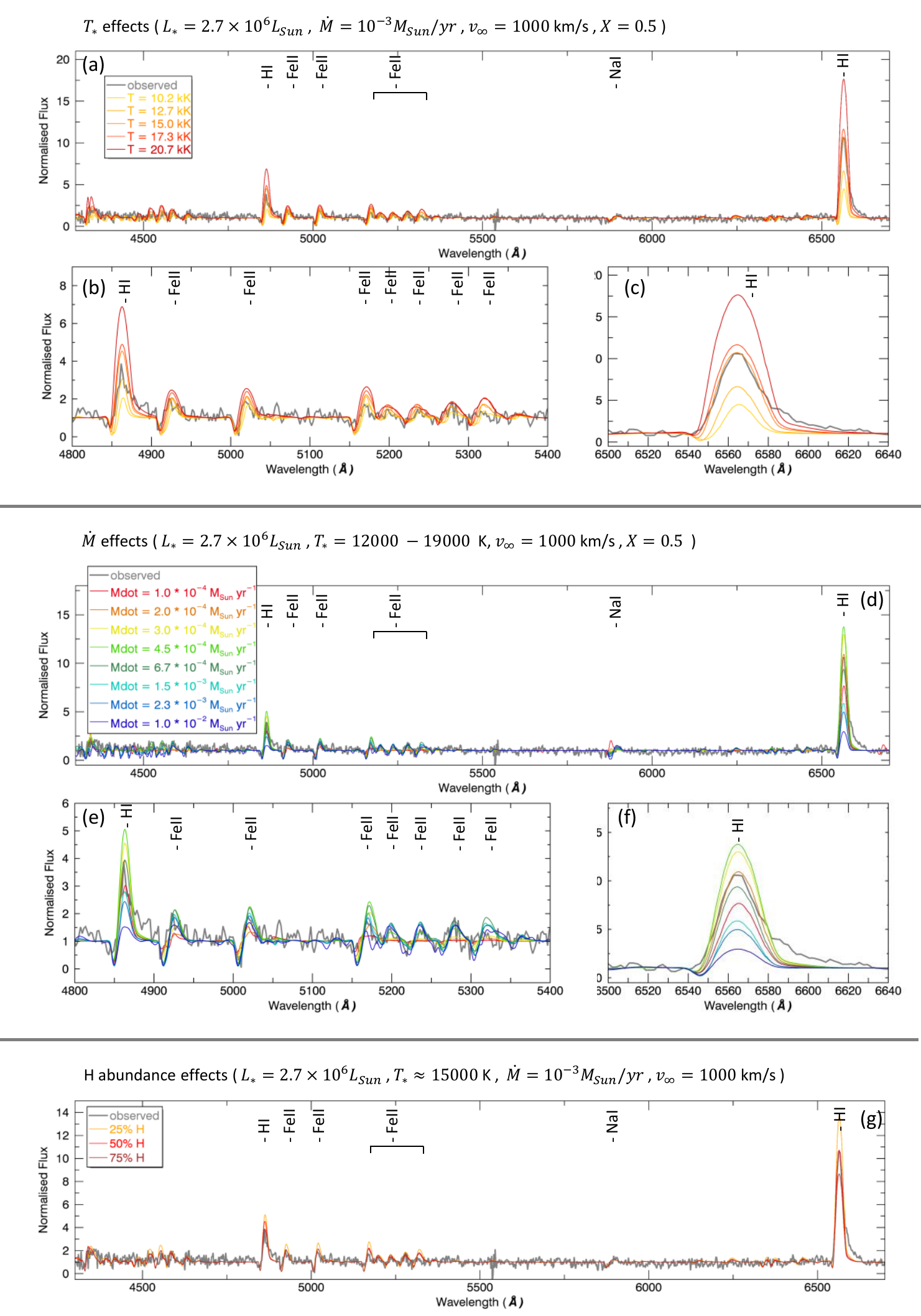}
\caption{Grid of CMFGEN synthetic spectra around the best-fit model of 2015bh's progenitor, exploring variations in \tstar\ (Panels $a-c$), \mdot ($d-f$), and H abundance ($g$). The model parameters are indicated above each group of panels.}
\label{specdegen}
\end{figure*}

-- {\it Temperature:} Figure~\ref{specdegen}a shows a set of models with $\lstar =2.7 \times 10^{6} ~ \lsun$, $\mdot = 10^{-3} ~\msunano$, $\vinf = 1000$ \kms, identical abundances (Table \ref{modelparam}), and different $\tstar$ in the range $10200 - 20700~\K$. Going from lower to higher temperatures, the emission of all \ion{H}{i} lines (see H$\alpha$ in Fig.~\ref{specdegen}c) and \ion{Fe}{ii} lines (Fig.~\ref{specdegen}) increases. This happens because an increase in $\tstar$ leads to more H ionization, up to the point where \ion{H}{} is essentially fully ionised in the region of the CSM where H$\alpha$ originates. The only line not affected by the effective temperature is \ion{Na}{i} $ \lambda 5889 ~ \ang$, due to the fact that \ion{Na}{} is completely neutral in this temperature range.

-- {\it Mass-loss rate:} Another factor that shapes the ionization structure is the mass-loss rate. Figure~\ref{specdegen}d contains a set of models with fixed $\lstar$, $\vinf$ and abundances (Table~\ref{modelparam}), but exploring variations in $\mdot$ from $10^{-4} ~\msunano$ to $10^{-2} ~\msunano$. A model with $\mdot = 10^{-4}~\msunano$ (red line in Fig.~\ref{specdegen}d) underestimates all emission lines compared to the observed spectrum. Increasing $\mdot$ by a factor of 2 (orange line in Fig.~\ref{specdegen}d) produces a better fit of the H$\alpha$ line (Fig.~\ref{specdegen}f), but grossly underestimates \ion{Fe}{ii} $ \lambda \lambda 5198, 5235, 5276, 5317 ~\ang $ (Fig.~\ref{specdegen}e). At $\mdot \geq 4.5 \times 10^{-4} ~\msunano$, the \ion{H}{i} emission lines decreases. For example, a model with $\mdot = 6.7 \times 10^{-4} ~\msunano$ shows a lower \ion{H}{i} emission than observed. A model with $\mdot$ somewhere in the range $4.5 \times 10^{-4}$ - $ 6.7 \times 10^{-4} ~\msunano$ would also fit the observed spectrum. Rising $\mdot$ to values higher than $6.7 \times 10^{-4} ~\msunano$ results in less \ion{H}{i} emission. This is caused by the recombination of ionised hydrogen into neutral hydrogen. Figure~\ref{ionstruct} shows that as $\mdot$ increases, more and more ionised hydrogen recombines. When we reach $\mdot = 6.7 \times 10^{-4} ~\msunano$, neutral hydrogen dominates in the formation region of $H \alpha$, leading to reduced emission. In these models the temperature varies slightly together with the mass loss rates, due to the method we followed in developing this grid. As previously explained, a rise in temperature would lead to a rise in the emission lines, therefore the increasing temperature in the lower mass-loss rate side only leads to a faster increase in ionisation, i.e. if we had a model with $\tstar = 10000$ K, we would need a much lower mass loss rate to find a fit for the \ion{H}{$\alpha$} emission. Models with $\mdot < 3 \times 10^{-4} ~\msunano$ do not fit the observed spectrum, since no satisfactory fit can be simultaneously obtained for \ion{}{H$\alpha$}, \ion{H}{$\beta$}, and \ion{Fe}{ii} lines in this regime. For $\mdot < 3 \times 10^{-4} ~\msunano$, the \ion{Fe}{ii} lines are too weak and an increase in temperature is needed to fit \ion{H}{$\alpha$} and \ion{H}{$\beta$}. However, this causes \ion{He}{i} $\lambda 5875.66 ~ \ang$ emission  on the red side of \ion{Na}{i} $\lambda 5889.95 ~ \ang$ in the synthetic spectrum, which does not fit the observations.

On the high $\mdot$ end, changes in $\mdot$ have a much greater impact on the models. For example, for $\mdot = 1.5 \times 10^{-3} ~\msunano$ we already require $\tstar \simeq 19500 ~\K$ to find a reasonable fit for the observations. For $ \mdot > 2.3 \times 10^{-3} ~\msunyr$ we could not fit the emission lines, since all H is already ionised. Therefore a further increase in temperature would not lead to H$\alpha$ emission comparable to the observations.

-- {\it Clumping:} Our models underestimate the observed red-wing emission of the \ion{H}{$\alpha$} and \ion{H}{$\beta$} lines (Fig. \ref{modelfit} ). Models with unclumped winds show increased electron scattering emission as expected \citep{hillier91}, however this is not sufficient to explain the feature observed in our spectrum (Fig. \ref{clump}). It is unclear to us what is the origin of this feature. One possibility is that the progenitor at this point has a time-dependent wind. Another possibility would be that the velocity structure is more complex than our assumptions and presents regions of higher wind velocities. If the outflow is indeed unclumped, then a higher $\mdot$ than that quoted in Table \ref{modelparam} is required to fit the observed spectrum. For e.g., the unclumped wind model shown in Fig. \ref{clump} has $\mdot = 6 \times 10^{-3} ~\msunano$. This is not proportional to $\frac{1}{\sqrt{f}}$ due to the change in the optical depth structure.      

-- {\it Velocity law:} As mentioned in Section 2, the $\beta$ parameter gives the steepness of the velocity law, thus changing the density structure in the inner regions. A decrease in $\beta$ would require an increase in $\mdot$ in order to keep $\rho$ constant. Our models show that a decrease to $\beta=1$, overestimates the emission lines. The spectrum fits well again once $\mdot$ is increased to $1.65 \times 10^{-3}$ \msunano. Given the relatively small change in $\mdot$ due to the change in the $\beta$ parameter, the variation of $\beta$ is covered in our determined range for $\mdot$. In addition, the shape of the emission line is slightly affected by the change in the velocity/density structure, and our $\beta=1$ model clearly shows a poorer fit to the line morphology.

\begin{figure}
\centering
\includegraphics[scale=0.53]{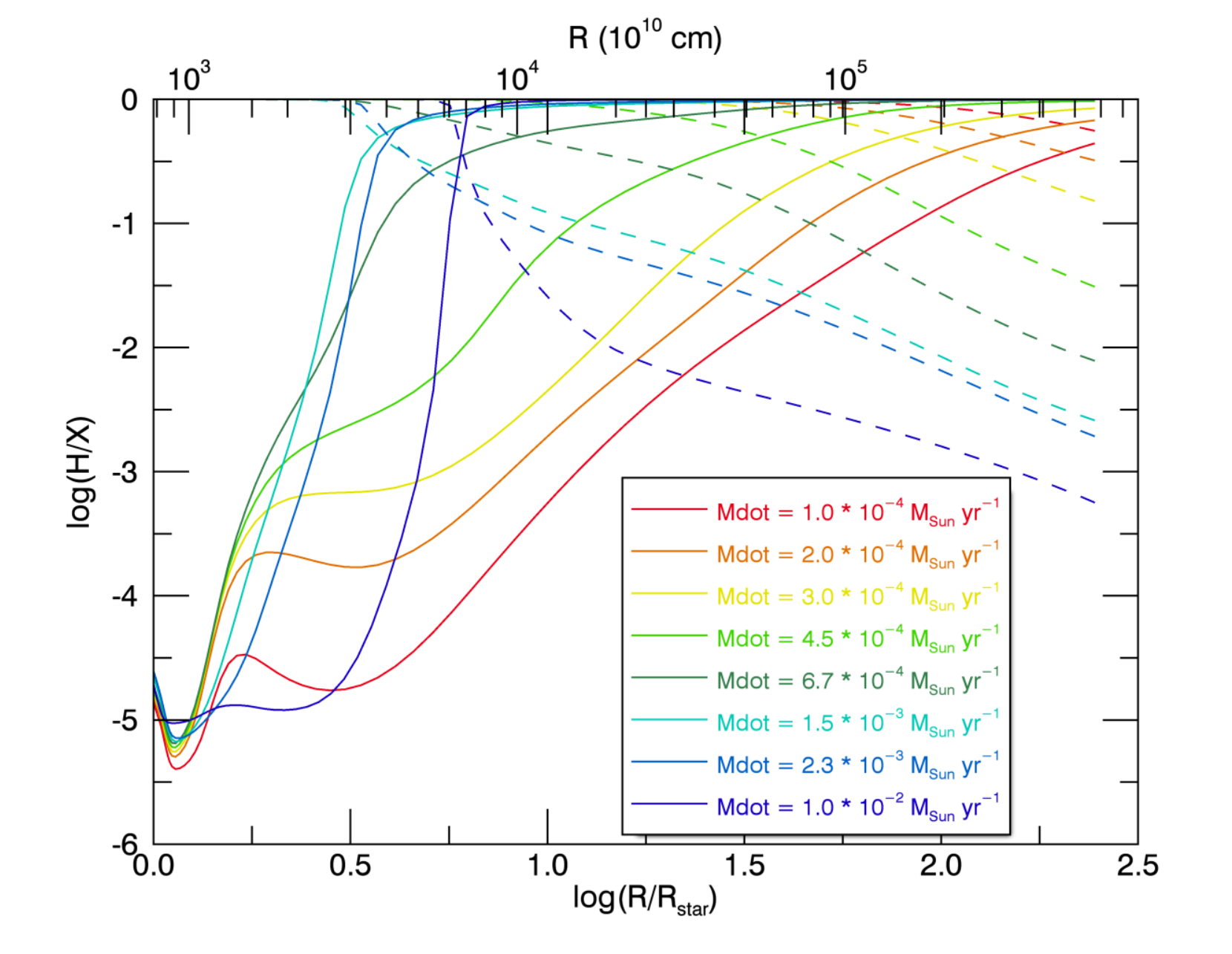}
\caption{H ionisation structure for SN2015bh progenitor models with different values of \mdot. The continuous line represents the amount of neutral hydrogen, while the dotted line represents the ionised hydrogen.}
\label{ionstruct}
\end{figure}

\begin{figure}
\centering
\includegraphics[scale=0.7]{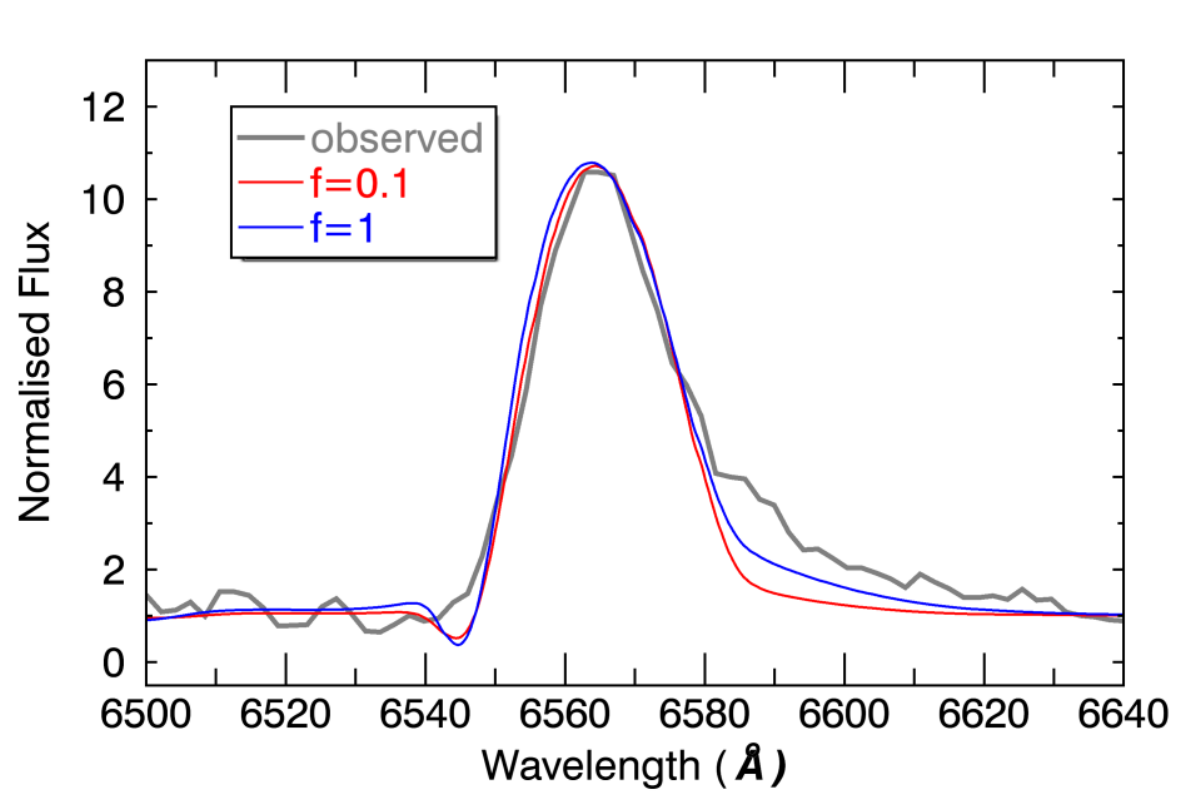}
\caption{Comparison of \ion{H}{$\alpha$} line between our best-fit model from Fig. \ref{modelfit} which has a clumped wind with $f=0.1$ (red) and a model with an unclumped wind, i.e. $f=1$ (blue). The observed spectrum is included in gray.  }
\label{clump}
\end{figure}

-- {\it Abundances:} Figure~\ref{specdegen}g shows the spectra of three identical models except for the abundances of \ion{H}{} and \ion{He}: the red spectrum is our best-fit model discussed in Section 3.1, having the abundances presented in Table \ref{modelparam} (i.e. $X=0.5$); the orange spectrum corresponds to a model having $X=0.25$ and the brown spectrum to a model with $X=0.75$. One would expect that an increase in the \ion{H}{} abundance would lead to stronger emission, but we observe the opposite effect, where the $X=0.25$ spectrum has a stronger \ion{H}{$\alpha$} emission line than any of the higher abundance models. The explanation is similar to the increased \mdot~ case. In this \tstar~ range, having more \ion{H}{} in the CSM leads to more recombination to neutral hydrogen, and therefore a decrease in the strength of the \ion{H}{} emission lines.

The \ion{Fe}{ii} $ \lambda \lambda 4924, 5018, 5169 ~ \ang$ lines also show increased strength for a lower \ion{H}{} content even if the \ion{Fe}{} abundance remains unchanged. Changes in the \ion{H}{} or the \ion{Fe}{} abundances will have an effect on both ionization structures, ultimately affecting the \ion{H}{} and \ion{Fe}{} line strengths. In addition, a change in the \ion{H}{} abundance also affects the temperature structure due to the different opacity of \ion{H}{} and \ion{He}{}. Taking into account the effect of the temperature on the \ion{H}{} emission, a decrease in \tstar~ to only $14000 ~\K$ would lead to a well-fitted spectrum for the model with $X=0.25$. Similarly, a small increase in \tstar~ would lead to a good reproduction of the observed spectrum for a model with $ X=0.75$. This means that we have a poor constraint on the abundances based solely on this observed spectrum.

We also computed a model with $X=0.05$ to investigate if extremely low values of the \ion{H}{} abundance would fit the observed spectrum. We found a poor fit that underestimates the \ion{H}{$\alpha$} and \ion{H}{$\beta$} emission and overestimates all \ion{Fe}{ii} emission lines. Increasing the mass loss rate to $ \mdot = 3.4 \times 10^{-3} ~\msunano$ and decreasing the \ion{Fe}{} abundance to a quarter of the solar \ion{Fe}{} metallicity, we found a fit for most emission lines in the optical spectrum. However, the absorption component of the \ion{Na}{i} $ \lambda 5889 ~\ang$ is now filled by the \ion{He}{i}$\lambda 5876 ~\ang$ emission line, which is not observed. This is a good indication that we have overestimated the \ion{He}{} abundance and, by consequence, underestimated the amount of \ion{H}{}.

To conclude, we found that there are a number of combinations of model parameters that fit the observed spectrum. The mass loss rate - temperature values are degenerate, however not as much as to significantly affect the conclusions on the properties of the SN2015bh progenitor. We can confidently place our mass loss rate in the $6 \times 10^{-4} ~\msunano $ to $ 1.5 \times 10^{-3} ~\msunano$ interval and the temperature in between $13000 ~\K$ and $19500 ~\K$. The abundances of \ion{H}{} and \ion{He}{} are extremely difficult to determine by modelling the 12 November 2013 spectrum alone. We suggest that the \ion{H}{} and \ion{He}{} abundances could be better constrained by modelling the post-explosion spectra that shows simultaneously the presence of \ion{H}{}, \ion{He}{i}, and \ion{He}{ii} lines. 

\section{Discussion}

\subsection{Constraints on the Exploding Star: a warm LBV}
Our CMFGEN models of the 12 November 2013 spectrum of SN 2015bh suggests that the progenitor had $\lstar =1.8 -4.7  \times 10^6 ~ \lsun$, $\tstar = 13000 - 19500 ~\K$, $\teff = 8700-10000 ~\K$, $ \mdot = 0.67
 -1.5 \times 10^{-3} ~ \msunano$, $\vinf = 1000~\kms$, $X=0.25-0.75$, and half-solar \ion{Fe}{} abundance. We interpret the spectrum as arising from the extended photosphere and stellar wind, similar to other observed LBVs in the Galaxy and SMC. We do not see any obvious evidence for interaction in 12 November 2013. Let us now compare SN2015bh's progenitor properties with those of different classes of evolved massive stars. 

The spectral morphology of the progenitor of SN2015bh strongly resembles an LBV. The luminosity computed in our model of a few times $10^{6} ~\lsun$ corresponds to typical LBV luminosities ($\sim 10^{5} - 10^{6.7} ~\lsun$; \citealt{vg01,smithvink04,clark09,clark05,gmg13}). The temperature range of our model lies in the mid-range of LBV temperatures ($8000 - 25000 ~\K$; \citealt{vg01}). LBVs have a large range of possible mass-loss rates stemming from quiescent stellar winds or eruptions, from $10^{-5} ~\msunano$ to $1 ~\msunano$ \citep{smith14araa}. Our determined value of \mdot\ fits well within this range. Our models indicate $\vinf=1000 ~\kms$, which is higher than the velocities estimated for LBVs in the Milky Way and Magellanic Clouds \citep{smithvink04}. One of the highest observed velocities of an LBV outflow was during the Eta Carinae Great Eruption ($\vinf = 600 - 800 ~\kms$;  \citealt{smith06}). Interestingly, \cite{izotov09} detected LBVs with fast winds ($800 ~\kms$) in low-metallicity dwarf galaxies, which are more in line with our derived \vinf~ for SN2015bh's progenitor. Our derived value of \mdot\ is on the high side for LBVs in quiescence, being similar to that of the current wind of Eta Car \citep{ghm12,hillier01}. While radiation pressure on lines and continuum could drive the winds of Eta Car \citep{hillier01} and other bona-fide LBVs such as AG Car \citep{ghd09,ghd11}, we cannot exclude that the star possess a dynamic super-Eddington wind \citep{shaviv01,owocki04,vanmarle09}, especially in epochs when $M_\mathrm{R} \lesssim -12$~mag. Note that if the progenitor of SN 2015bh is in a super-Eddington state our previously derived $M_\mathrm{min}$ is not applicable.

Our results reinforce the suggestions from previous studies that the progenitor of SN2015bh is an LBV. \cite{thone16} propose that the pre-explosion spectrum is very similar to that of a quiescent LBV based on the combined stellar evolution and atmospheric models from \citet{gme14}. \cite{goranskij16} support the LBV progenitor scenario based on the argument that the brightness measurements from 6 March 2008 are close to the Humphreys \& Davidson limit. \cite{eliasrosa16} also conclude that the progenitor of SN2015bh was likely a massive blue star. 

The high \vinf\ could instead be indicative of a Wolf-Rayet (WR) progenitor. However, the derived value of \teff\ for SN2015bh's progenitor is significantly below typical values for the \teff\ of WRs ($30000-150000~\K$; \citealt{crowther07,sander12}). A more plausible possibility is that the progenitor of SN2015bh was a WR star that inflated just before eruption, i.e. a star with an increased envelope radius due to the star's proximity to the Eddington limit. This would in principle decrease the effective temperature. \cite{grafener12a} showed through numerical models that the effective temperature of a H-poor WR ($X<0.05$) can be reduced from $100000~\K$ to $40000~\K$ through envelope inflation, while LBVs of solar abundances can have \teff\ as low as $16000~\K$ when inflation takes place. While none of the models presented in \cite{grafener12a} match our SN2015bh properties, we cannot exclude the possibility of an inflated WR progenitor, especially since our CMFGEN model is very close to the Eddington limit in the deep atmospheric layers ($\Gamma \sim 0.85 - 0.90$). 

We can exclude a red supergiant (RSG) or yellow hypergiant (YHG) progenitor for SN2015bh, since RSGs have temperatures of at most $ 6000 ~\K$ \citep{levesque05,davies13} and YHGs are in the range of $4000 $ to $8000 ~\K$ \citep{kovtyukh07} and the wind velocities of these types of stars are typically up to $100 ~\kms$ (\citealt{dejager98}). \cite{thone16} discusses the possibility of a YHG progenitor based on the photometry and on the assumption that prior to 2013, the star was quiescent. While the authors raise it as a viable option, they question the lack of dust surrounding the progenitor, which would be expected in stars at low temperature, and the high \vinf.

Even if \teff, \lstar\ and \vinf\ are consistent with those of supergiant B\big[e\big] stars (sgB\big[e\big]), which typically have $\teff \sim 10000-25000~\K$ and $\lstar > 10^4 ~\lsun$ \citep{lamers98br}, the absence of forbidden lines are a clear indicator that the progenitor of SN2015bh was not a B\big[e\big] star. Furthermore, sgB\big[e\big] stars generally do not exhibit the large photometric variations seen in the progenitor of SN2015bh, and the lack of an infrared excess is also at odds with a B\big[e\big] classification.

\subsection{The Progenitor of SN2015bh as an Interacting Binary?}
Since a significant fraction of massive stars are found in binary systems \citep{sana12,sana17}, we now explore the possibility that SN2015bh's progenitor had a companion star. This scenario is attractive especially because the LBV-like properties of the progenitor of SN 2015bh (Sect. \ref{res.properties}) before the 2015 events bare significant resemblance to those of HD 5980, in particular the \lstar, \tstar, \mdot\ and \vinf. In addition, the similarity in the light curve variation pattern and time-scale raises the question of whether SN 2015bh's progenitor was a similar system as HD 5980 but later underwent a powerful, and perhaps terminal, explosion.  

HD 5980 is a massive multiple-star system in the Small Magellanic Cloud \citep{koenig06}. It is formed by an eclipsing pair of stars, stars A and B, and star C, which is itself a binary \citep{naze18}. The system shows long-term variability in the light curve for $\sim 40$ years and sudden variability in 1993 and 1994. The cause of the long-term variation is believed to be due to one of the stars in the system, star A, an LBV undergoing S-Doradus variability. The properties of star A change rapidly, especially during the 1994 event, having $\vinf=500-2440 ~\kms$, unclumped $\mdot=10^{-5}-10^{-4} ~\msunano$, $\teff=23000-43000 ~\K$, $\log(L/\lsun)=6.3-7.05$ \citep{georgiev11}.

\citet{grafener12a} suggests that star A in HD5980 could have been an inflated LBV during its 1994 event due to the proximity to the Eddington limit. The increase in radius could trigger an eruption, in a similar fashion as proposed by \citet{smith11} to explain the 1840's Eta Carinae eruption. For 2015bh, the sudden increases in absolute magnitude in the 2008 and 2013 events could also be triggered by envelope inflation followed by binary interaction, especially because our models indicate that the star is likely close to the Eddington limit.

A binary scenario has also been advocated in the context of the pre-explosion outbursts of SN 2009ip \citep{mauerhan13} and SN 2015bh \citep{soker16}. If SN 2015bh's progenitor is in a binary system, the secondary star would have to be much fainter than the primary, since we do not see any evidence of spectral lines from the companion in the optical spectrum from 12 November 2013. A higher cadence over a long period of time would be required to obtain orbital information of the putative binary system. Future high-cadence surveys such as LSST could possibly observe the progenitors of similar events as SN2015bh and establish their long-term photometric behaviour with unprecedented detail.

\subsection{Implications and Comparison to Other LBVs and Interacting Supernovae}

\begin{figure}
\includegraphics[width=1.1\columnwidth]{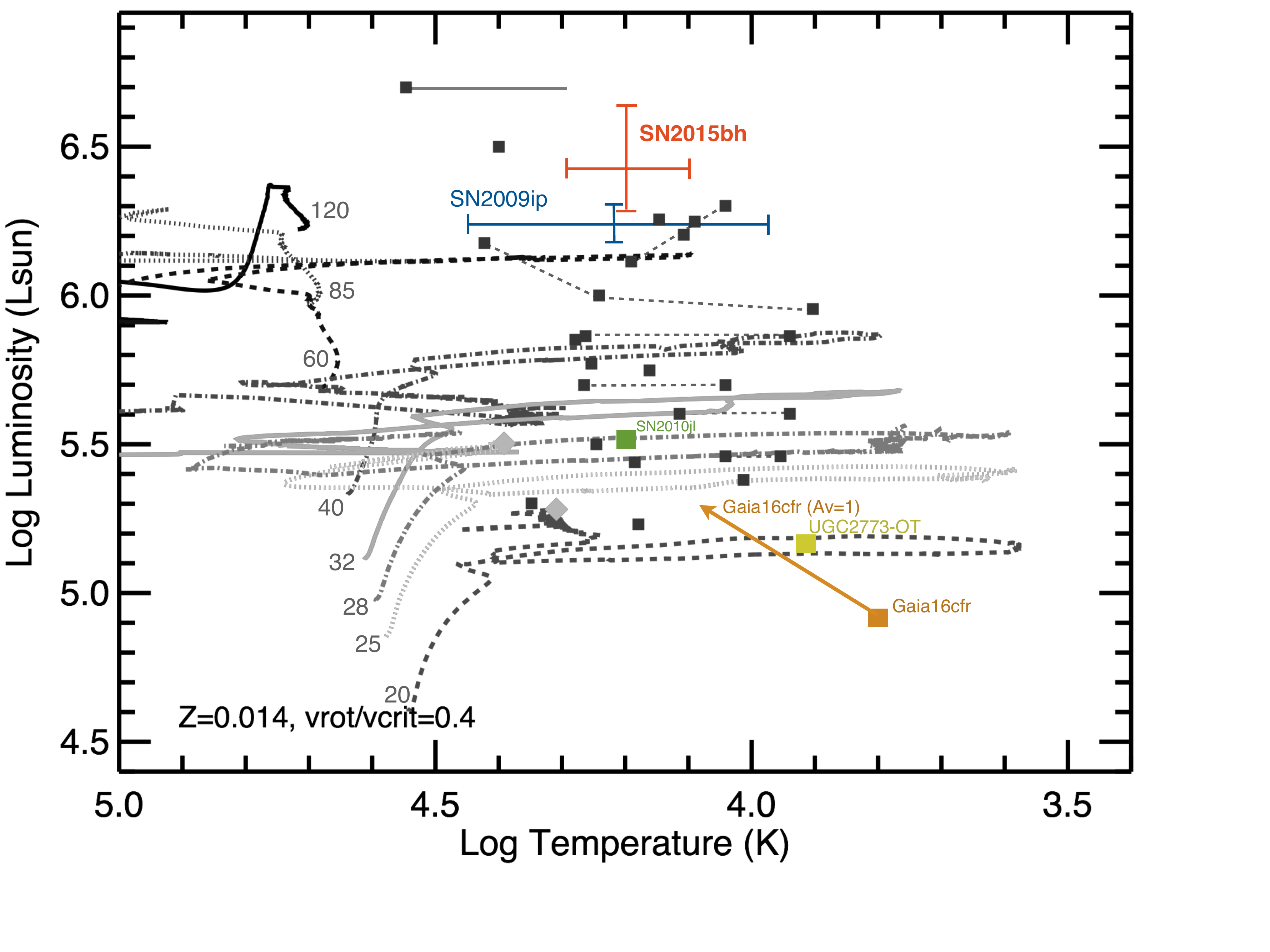}
\caption{HR diagram showing the location of the progenitor of SN2015bh as determined in this work (red). We also include the progenitor of SN2009ip (blue; \citealt{smith10_sn2009ip,foley11}), SN2010jl (green; \citealt{smith11_sn2010jl}), UGC2773-OT (yellow; \citealt{smith10_sn2009ip}), and Gaia16cfr (orange; \citealt{kilpatrick17}). We also show the location of Galactic LBVs in quiescence (black squares; \citealt{gmg13} and references therein) and Geneva evolutionary tracks for single stars at solar metallicity for reference \citep{ekstrom12,georgy12a,gmg13}. }
\label{hrd}
\end{figure}

While it was originally thought that LBVs cannot be direct progenitors of SN and would instead always evolve to WR stars, recent observations and modelling have strongly suggested otherwise. Figure~\ref{hrd} shows the location of SN2015bh's progenitor in the HR diagram, together with those of galactic LBVs and interacting SNe with similar light curves for which determinations of $L$ and $\teff$ exist in the literature. The temperature range of SN2015bh's progenitor compares well to other LBVs, while the luminosity places our progenitor in the high end of LBV luminosities. While the temperature of SN2009ip's progenitor is poorly constrained, our results indicate that the progenitor of SN2015bh is slightly more luminous than the progenitor of SN2009ip, despite the fact that their light curves are extremely similar. The placement in the HR diagram of the progenitor of SN 2015bh also points to an initial mass of $M \simeq 150-200 ~\msun$.

\begin{figure}
\centering
\includegraphics[width=1.02\columnwidth]{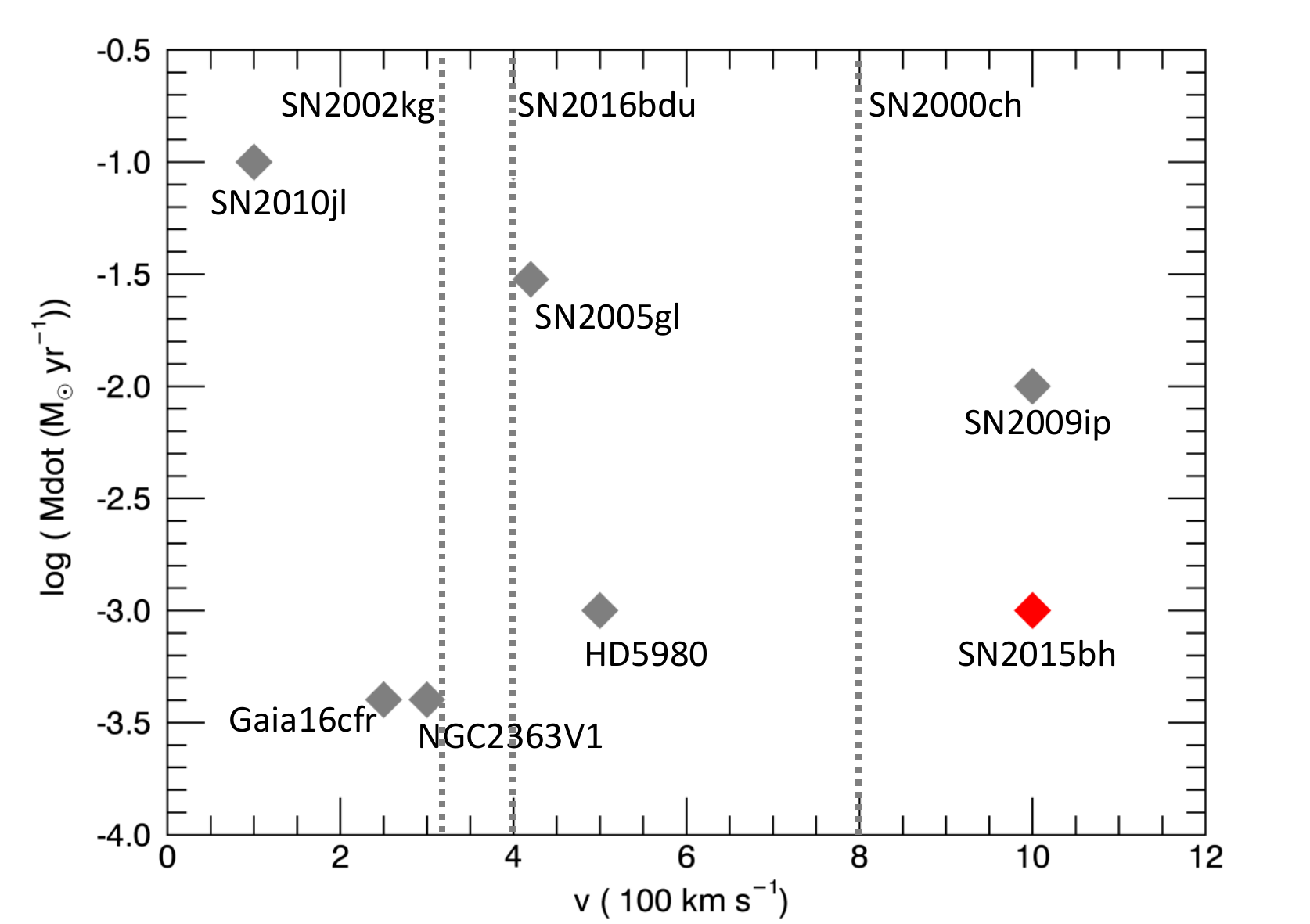}
\caption{Comparison between the wind/CSM velocities and mass-loss rates of interacting SNe similar to SN2015bh. The values for SN2015bh (red) are taken from this work, and the references  for the other events are as follows:  SN2005gl - \cite{galyam09}, SN2010jl - \cite{fransson14,dessart15}, SN2009ip - \cite{ofek13_2009ip}, SN2002kg - \cite{vandyk06}, SN2000ch - \cite{wagner04}, HD5890  - \cite{georgiev11},  NGC2363v1 - \cite{drissen01}, Gaia16cfr - \cite{kilpatrick17}, and SN2016bdu - \cite{pastorello17}.}
\label{mdotvel}
\end{figure}

Significantly more information exists about the progenitor \mdot\ and \vinf\ of SNe with light curves similar to SN 2015bh. Figure~\ref{mdotvel} compares the \mdot\ and \vinf\ derived for both terminal and non-terminal events. We can see that while they all share values of \mdot\ and \vinf\ characteristic of LBVs, the range of values likely indicate that these events do not have exactly the same progenitor, but rather span a range of masses and luminosities. 

\begin{itemize}
\item {\it SN2005gl} is believed to have a $ \mstar > 50~ \msun$ LBV progenitor (\citealt{galyam07}) that ejected an outflow with $\mdot = 0.03~\msunano$ and $ \vinf \simeq 420~\kms$ (\citealt{galyam09});
\item {\it SN2010jl} likely originated from a $>30 ~\msun$ star (\citealt{smith11_sn2010jl}) that had a high \mdot\ of $ 0.1 ~\msunano$ before exploding (\citealt{dessart15,fransson14}); 
\item {\it SN2009ip} shows almost identical photometry to SN2015bh and signs of LBV-like behaviour \citep{thone16}; the mass-loss rate of SN2009ip's progenitor was determined by \cite{ofek13_2009ip} to be $10^{-3}$ to $10^{-2} ~\msunano$ and $\vinf \simeq 1000~\kms$, while \cite{moriya15} suggested an LBV progenitor with $\mdot = 0.1 ~\msunano$ and $\vinf = 550 ~\kms$;
\item {\it SN2002kg} has a lower velocity of $\vinf=320~\kms$ \citep{vandyk06} and \mdot\ is unconstrained;
\item {\it SN2000ch} displays the highest wind velocity in this sample ($800~\kms$; \citealt{wagner04}), which is similar to what we found for SN2015bh's progenitor;
\item {\it SNhunt248} is another event exhibiting a very similar behaviour to SN2015bh, in light curve, spectra and the lack of dust in its close environment; however, SNhunt248's spectrum suggest a lower wind velocity of $<300~\kms$ \citep{kankare15}.
\end{itemize}

\begin{figure*}
\centering
\includegraphics[scale=0.6]{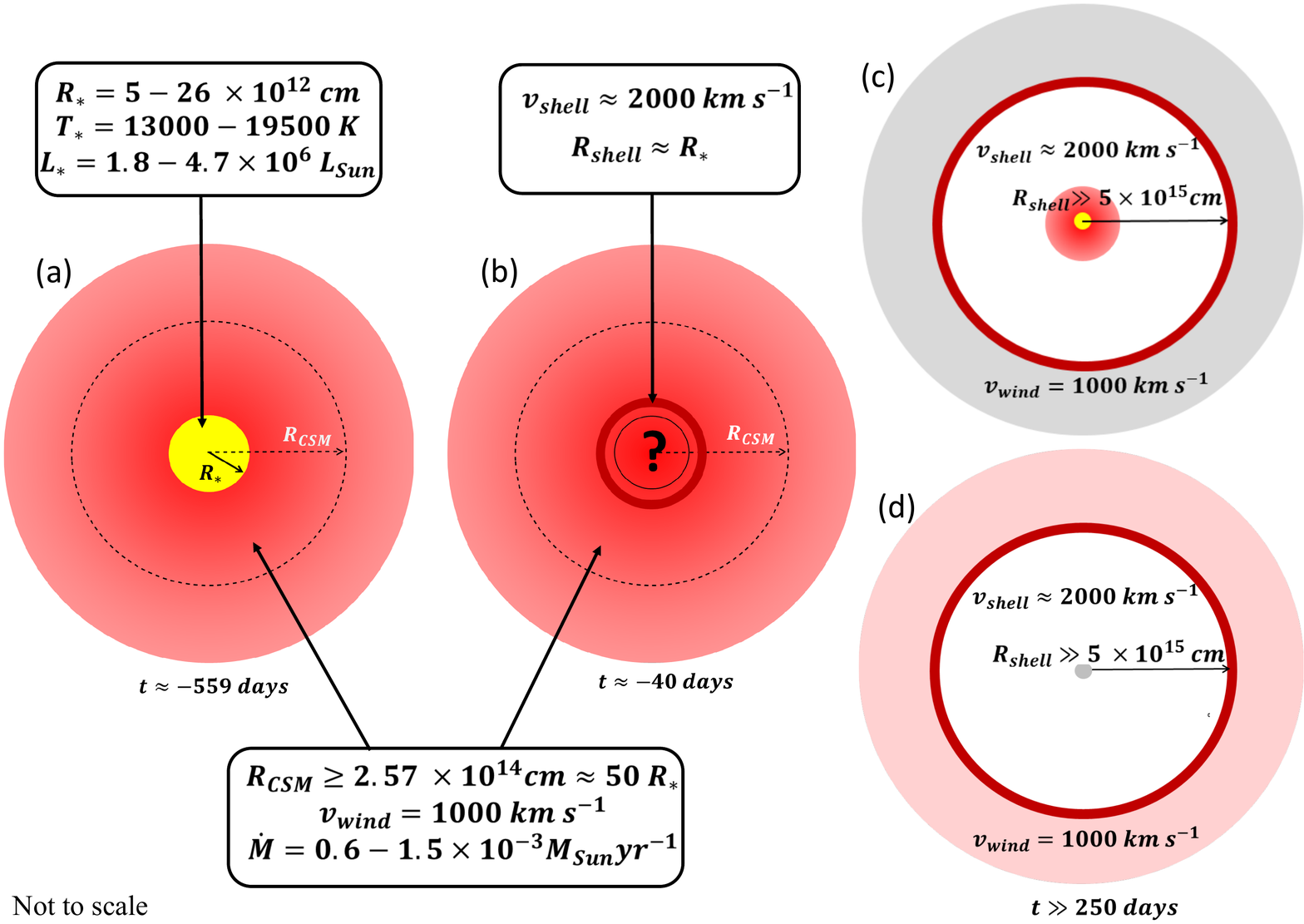}
\caption{Sketch of the CSM geometry around SN2015bh at selected epochs. Panel (a) shows the pre-explosion extended CSM that we derived in this paper, (b) shows the expansion of the ejected shell into the extended CSM assuming the shell is ejected at $t \simeq -40$ days, hence the radius of the shell $R_\mathrm{shell}=\rstar$ at this stage, (c) shows a post-explosion scenario where the star survived, and (d) shows a post-explosion scenario for a supernova explosion.}
\label{csmscen}
\end{figure*}

\subsection{Constraints on the Post-Explosion Behaviour}
Having previously discussed the properties of the SN2015bh progenitor, let us examine the possible scenarios for the post-explosion behaviour of SN2015bh in light of our new findings. We follow the definition of $t=0$ as 24 May 2015, i.\,e. at the peak of the 2015B event as in \citet{thone16}. Thus, the 12 November 2013 spectrum was taken at $t = -558.8~\mathrm{d}$ and the peak of the 2015A event was at $t \simeq -25~\mathrm{d}$.

We begin this subsection by briefly summarizing the scenarios that have been proposed in the literature.

{\it - Interaction of two shells + dense CSM \citep{thone16}}: in this scenario, a first thin shell with $v=700~\kms$ would be ejected years prior to 2015 B, giving rise to the absorption component observed at $t = -558.8~\mathrm{d}$ (see their Fig. 13). Then, a second shell with $ \simeq 2200-2300 ~\kms$ would be ejected at the time of the 2015 A event, which could be the ejecta of a genuine SN or SN impostor. The second shell would appear first in absorption and, about a 100 days later, catch up with the first shell and produce the $2000 ~\kms$ emission component by shock excitation. These authors also proposed that at late times the $2300\,\kms$ shell interacts with a dense CSM that had been previously created by the progenitor. The \citealt{thone16} model reproduces the late evolution of the lightcurve with a shell of mass $M_s=\sim0.005-0.5~\msun$ and $v_s=2300 ~\kms$ interacting with a CSM created by a progenitor mass-loss rate of $\mdot\simeq0.005~\msunyr$. This is 3 to 10 times higher than the \mdot\ we determined from our CMFGEN models. Radiative hydrodynamic models can be used to infer the explosion energy, mass, energy released in the wind and the expansion velocity of the shocked shell given our determined wind properties and the lightcurve evolution \citep{moriya15}, for a more detailed comparison to the existing literature.

{\it - SN ejecta + fallback + CSM interaction with multiple shells \citep{eliasrosa16}}: these authors propose that a $\sim1000\,\kms$ shell was expelled around 2002 or before to explain the pre-explosion spectrum. They conclude that an outburst at the end of 2013 with material moving at $1000$ $\kms$ has produced the increase in luminosity observed in December 2013. They propose that the 2015 A event was a core-collapse SN that experienced large fallback of material, thus explaining the low energy. They also infer the presence of three emission components in \ion{H}{$\alpha$}, which would be an argument pro core-collapse as a multicomponent \ion{H}{$\alpha$} is typical of interacting SNe. Then at $t=0$, the newly ejected material collides with a slow, dense, CSM and produces 2015 B. After the shock passes, the two absorption components ($1000 ~\kms$ and $2100 ~\kms$) are visible again. They attribute this to at least two shells or clumps of cooler material moving along the observer's direction.

Our findings imply that the pre-explosion spectrum from $t = -558.8~\mathrm{d}$ is formed in an extended CSM, unlike  \citet{thone16} and \citet{eliasrosa16} who had proposed that the pre-explosion spectrum was formed in a thin shell. We propose that the subsequent interaction occurs mainly with the extended material, eliminating the necessity of the first shell ejection that has been proposed in \citet{thone16}. Figure \ref{csmscen}a shows our proposed geometry at $t = -558.8~\mathrm{d}$ based on our derived values of $\rstar=5.1-25.9 \times 10^{12}$ cm, $\tstar=13000-19000~\K$, $\lstar=1.8-4.7 \times 10^{6} ~\lsun$, $\mdot=0.6-1.5\times10^{-3} ~\msunano$, and \vinf=$1000~\kms$. Given that the progenitor star had similar high luminosities for at least 20 years before the eruption, during this period the star was most likely blowing a wind at least as dense as the one we infer for $t = -558.8~d$. This long-term wind produces an extended CSM with $R_\mathrm{CSM} \gtrsim 2.57 \times 10^{14}$ cm as determined in Sect. 3.2. The wind needs to be ejected at least $30$ days prior to the 12 November 2013 observations, i.e. $t= -589~\mathrm{d}$, in order to produce the observed emission. While our stationary models yield an outflow with $\rho \propto r^{-2}$ at large distances, the extended pre-explosion CSM around SN 2015bh is most likely inhomogeneous given the significant variability before explosion \citep{thone16,eliasrosa16,ofek16}.

We support the suggestion from \citet{thone16} that an explosion (not necessarily terminal) occurs around $t \simeq - 40$ days, when a second absorption component at $-2000~\kms$ is observed in the \ion{H}{$\alpha$} line profile. However, we propose that the double absorption line profile is generated by a $2000 ~\kms$ shell expanding into an extended CSM produced by a long-term $1000 ~\kms$ wind. Figure \ref{csmscen}b sketches how the CSM would look at this stage according to our scenario (see Fig. 13 of \citet{thone16} for comparison).

At this point, it is uncertain whether the star has survived. At about $t=+26~\mathrm{d}$, the spectrum of SN 2015bh clearly shows two absorption components at $\sim -1000~\kms$ and $-2000 ~\kms$ and by  $t=100~\mathrm{d}$, \ion{H}{$\alpha$} has the two components in emission \citep{thone16,eliasrosa16}. These authors also reported that at $t=+242$ days, the two-peak emission is still present and in our scenario the fast shell would be now at $5 \times 10^{15} $ cm. As the shell expands, it will eventually become optically thin at very late times ($t>>250$ days). Below, we analyze two scenarios for $t>>250$ days depending on whether the star survived or collapsed.

{\it Scenario 1 (surviving star):} Lets first explore the scenario in which the star is still present after the main event, as originally proposed by \citet{thone16}. At first the star should be out of thermal equilibrium, and then re-establish a stellar wind. For $t>>250$ days,  we should see both emission from the optically-thin shell and from the new wind/CSM of the star inside the shell (Fig.~\ref{csmscen}c). If this is the case, then late time spectroscopy should reveal the presence of the new wind and CSM. The star could slowly relax back to its original LBV state or become Wolf-Rayet star, depending how the mass of the \ion{H}{} envelope before the explosion compares to the mass ejected in the process ($\sim0.5~\msun$; \citealt{thone16}). Curiously, the $+581$ days spectrum of SN2009ip (\citealt{thone16}) bares significant resemblance to the SN2015bh spectrum from 12 November 2013, showing prominent \ion{H}{$\alpha$} and \ion{H}{$\beta$} lines, multiple \ion{Fe}{ii} lines and the \ion{Na}{i} $ \lambda 5889 \ang$ with a strong absorption component. Our scenario of a surviving star is similar to that proposed by \citet{thone16}, with the exception that we consider only one high-velocity, optically-thin shell expanding into an extended CSM created by the progenitor.

{\it Scenario 2:} If the star did not survive, the $1000~\kms$ component would have to come from the outside of the $2000~\kms$ shell. Since we propose that the $1000~\kms$ component arises in the extended progenitor wind, this component should become weak and eventually disappear for $t>>250$ days. There is the possibility that the shell is still interacting with the former CSM (Figure \ref{csmscen}d), which would further support our argument that the progenitor had an extended wind and not a thin shell prior to 2015A. If the star exploded as a genuine SN, 2015bh would be a remarkable case of a successful core collapse of a star of at least $35~\msun$ at the pre-SN stage.

\section{Summary and conclusions}

In this paper, we investigate the progenitor of SN2015bh by means of CMFGEN radiative transfer modelling of the spectrum obtained 559 days prior to the main event. Here we summarise our main findings: 

\begin{enumerate}
\item We perform radiative transfer modelling  of the SN2015bh progenitor using CMFGEN. Our models assume extended, clumped, steady, spherically symmetric wind in non-local thermodynamic equilibrium following a $\beta$ type velocity law and having the density structure
 $ \rho \propto r^{-2}$. 
 
\item  By running an extensive grid of models we conclude that the models are degenerate and a good fit could be reproduced by stars with slightly different physical properties. While \tstar\ and \mdot\ are constrained in relatively small intervals of $13000$ to $19500$ K corresponding to $\mdot = 6 \times 10^{-4} ~\msunano$ to $1.5 \times 10^{-3} ~\msunano$, the \ion{H}{} abundance shows a large range of values. We can fit the spectrum reasonably well for a model with $X=0.25$, $\tstar=14000 ~\K$, $\mdot=10^{-3} ~\msunano$ as well as for another model with $X=0.75$, $\tstar \simeq 15300 ~\K$, $\mdot=10^{-3} ~\msunano$ .

\item  For $\lstar = 1.8 \times 10^{6} ~\lsun$, $\rstar = 5.15 \times 10^{12}$ cm, $\tstar = 19500 ~\K$, $\mdot = 0.6 \times 10^{-3} ~\msunano$ and $\vinf = 1000 ~\kms$, the emission observed in the spectrum is generated in the CSM up to  $R_\mathrm{CSM} = 2.57 \times 10^{14}$ cm. The wind needs to have started at least $30$ days prior to the 12 November 2013 observations in order to fill out $R_\mathrm{CSM}$ at the determined \mdot\ and \vinf\. The mass of material contained in this region is $M_\mathrm{CSM} = 0.5 \times 10^{-4} ~\msun$. 

\item Based on the values of our best-fit models, we propose that the progenitor of SN2015bh is either a warm LBV or an inflated WR star of at least $35 ~\msun$. This type of progenitor matches other types of SNIIn/impostors progenitors. 

\item We conclude that the progenitor of SN2015bh had a strong extended wind which interacted with a faster shell moving at $2000~\kms$. We were not able to determine whether the star exploded as a SN or not, but we do propose two scenarios. In the case of a non-terminal eruption, the star could reinstate its previous wind or become a WR, and at very late times ($t>>250$ days) we should see emission from the $2000$ \kms\ shell, that has become optically thin, and the stellar wind/CSM. In the case of a SN, the shell could be interacting with the  extended CSM, even at $t>>250$ days, and possibly produce observable signatures. The level of late-time interaction will depend on both the CSM and the SN ejecta properties and a detailed hydrodynamics model of the entire light-curve would be needed to fully investigate this aspect. Similar conclusions have been reached by \cite{thone16} and \cite{eliasrosa16} with the main difference being that the previous works propose that the possible SN ejecta interacts with multiple thin shells, while our work strongly supports interaction with an extended CSM. The different scenarios investigated above can be tested with future observations of SN 2015bh at late times.
\end{enumerate}

SN2015bh is comparable to many other events, such as SN2009ip, SNhunt248, HD5980, etc. Due to the unique pre-explosion spectrometry we were able to constrain the properties of SN2015bh's progenitor and, by extension, add to the still incomplete picture of SN progenitors and SN impostors.  Our results contribute to our understanding of the members of this new growing class of events and the pre-supernova evolution of massive stars.

\section*{Acknowledgements}
I.B acknowledges funding from a Trinity College Postgraduate Award through the School of Physics, and J.H.G. acknowledges support from an Irish Research Council New Foundations Award 206086.14414 "Physics of Supernovae and Stars''. The authors are grateful to Ylva G\"otberg for making available her python scripts for CMFGEN, and thank C. Th\"one, T. Moriya, and J.S. Vink for discussions on interacting supernovae and SN2015bh. We would like to thank the other researchers in the Astrophysics group at Trinity College Dublin for helpful discussions. 
\bibliography{refs.bib}
\bibliographystyle{./bibtex/aa}

\end{document}